\documentclass[conference,dvipsnames]{IEEEtran}

\usepackage[utf8]{inputenc}
\usepackage{url}
\usepackage{colortbl}
\usepackage{balance}
\usepackage{xspace}
\usepackage{graphicx}
\usepackage{verbatim}
\usepackage{bold-extra}
\usepackage{amsmath}
\usepackage{multirow}
\usepackage{listings}
\usepackage{boxedminipage}
\usepackage{soul}
\usepackage{enumitem}
\usepackage[normalem]{ulem}
\setlist{nolistsep,leftmargin=.5cm}
\usepackage{booktabs}
\usepackage{tabularx}
\usepackage{svg}
\usepackage{calc}
\usepackage{float}
\usepackage{multirow}
\usepackage[normalem]{ulem}
\useunder{\uline}{\ul}{}
\usepackage{amsmath}
\usepackage[most]{tcolorbox}
\usepackage{caption}
\usepackage{microtype} 
\usepackage{cite}
\usepackage{amsmath,amssymb,amsfonts}
\usepackage{algorithmic}
\usepackage{graphicx}
\usepackage{textcomp}
\usepackage{xcolor}
\usepackage[hidelinks]{hyperref}
\usepackage{ifthen}
\usepackage{wrapfig}
\usepackage{subcaption}
\usepackage{cleveref} 

{

\newcommand{\ie}{\textit{i.e.},\xspace}
\newcommand{\eg}{\textit{e.g.},\xspace}
\newcommand{\etc}{\textit{etc.}\xspace}
\newcommand{\etal}{\textit{et al.}\xspace}
\newcommand{\aka}{\textit{a.k.a.}\xspace}

\newcommand{\tool}{{\sc{Burt}}\xspace}
\newcommand{\app}{\tool}
\newcommand{\ap}{\tool}
\newcommand{\burt}{\tool}

\newcommand{\itr}{{\sc{Itrac}}\xspace}
\newcommand{\itrac}{\itr}

\newcommand*{\img}[1]{%
    \raisebox{-.2\baselineskip}{%
        \includegraphics[
        height=\baselineskip,
        width=\baselineskip,
        keepaspectratio,
        ]{#1}%
    }%
}

\usepackage{tikz}

\definecolor{bug_red}{rgb}{.84,.23,.29}

\definecolor{info-needed-color}{rgb}{1,.8,.12}

\newcommand{\crashscope}{{\scshape{CrashScope}}\xspace}
\newcommand{\tracereplayer}{{\scshape{TraceReplayer}}\xspace}
\newcommand{\avt}{{\scshape{Avt}}\xspace}

\newcommand{\fusion}{{\scshape{Fusion}}\xspace}
\newcommand{\ebug}{{\scshape{EBug}}\xspace}
\newcommand{\bee}{{\scshape{Bee}}\xspace}
\newcommand{\BugListener}{{\scshape{BugListener}}\xspace}

\def\BibTeX{{\rm B\kern-.05em{\sc i\kern-.025em b}\kern-.08em
    T\kern-.1667em\lower.7ex\hbox{E}\kern-.125emX}}

\begin{document}

\title{\tool: A Chatbot for Interactive Bug Reporting}

\author{
	\IEEEauthorblockN{Yang Song\IEEEauthorrefmark{1}, Junayed Mahmud\IEEEauthorrefmark{2}, Nadeeshan De Silva\IEEEauthorrefmark{1},
    Ying Zhou\IEEEauthorrefmark{3},\\
    Oscar Chaparro\IEEEauthorrefmark{1}, Kevin Moran\IEEEauthorrefmark{2}, Andrian Marcus\IEEEauthorrefmark{3}, Denys Poshyvanyk\IEEEauthorrefmark{1}}
	
    \IEEEauthorblockA{\IEEEauthorrefmark{1}\textit{William \& Mary}
    (USA), 
    \IEEEauthorrefmark{2}\textit{George Mason University}
    (USA), 
    \IEEEauthorrefmark{3}\textit{The University of Texas at Dallas}
    (USA)
    }
	\and
}

\maketitle

\begin{abstract}
This paper introduces \tool, a web-based chatbot for interactive reporting of Android app bugs. 
\tool is designed to assist Android app end-users in reporting high-quality defect information using an interactive interface. 
\tool guides the users in reporting essential bug report elements, \ie the observed behavior, expected behavior, and the steps to reproduce the bug.
It verifies the quality of the text written by the user and provides instant feedback. 
In addition, \tool provides graphical suggestions that the users can choose as alternatives to textual descriptions. 

We empirically evaluated \tool, asking end-users to report bugs from six Android apps. 
The reporters found that \tool's guidance and automated suggestions and clarifications are useful and \tool is easy to use. 
\tool is an open-source tool, available at \textit{\url{github.com/sea-lab-wm/burt/tree/tool-demo}}. 

A video showing the full capabilities of \tool can be found at \textit{\url{https://youtu.be/SyfOXpHYGRo}}. 

\end{abstract}


\section{Introduction}

Bug reports are essential for successful software maintenance and evolution. 
These reports are expected to provide clear and detailed information related to a defect, including the system's observed behavior (\textbf{OB}), the expected  behavior (\textbf{EB}), and the steps to reproduce the bug (\textbf{S2Rs})\cite{Bettenburg2008a}. 
Unfortunately, bug reports are often unclear, incomplete, and/or ambiguous -- often causing delays during the bug resolution process~\cite{GitHub2016,Zimmermann2012}. 

One main challenge that contributes to low-quality bug reports is the \textit{knowledge gap} between what reporters \textit{know} and what developers \textit{need}~\cite{Moran2015,Bettenburg2008a}. 
This is especially true when the reporters are software end-users, who are often unfamiliar with the system internals and do not know the information elements important for developers (\eg the OB, EB, and S2Rs) and how to express them.
Most bug reporting systems (\eg GitHub Issues, Jira, or Bugzilla) are not designed to address this knowledge gap, since they are static web forms and templates that: (1) offer limited guidance on \textit{what} information needs to be reported, and \textit{how} it needs to be reported; and (2) do not provide \textit{feedback} to end-users on whether the information they provide is clear and complete. 
In consequence, the burden of providing high-quality bug information rests mainly on the reporters. 
\looseness=-1

We propose a web-based chatbot for interactive BUg repoRTing (or \tool).
The \textit{software engineering challenge} addressed by \tool is ensuring high-quality bug reporting by end-users, considering the above-mentioned limitations of existing bug reporting systems. 
The \textit{envisioned users} for \tool are \textit{end-users} who report problems with their app. 
\tool guides the users during reporting essential bug report elements (\ie OB, EB, and S2Rs), offering instant quality verification, corrections, and graphical suggestions. 
\tool's \textit{usage methodology} is described in detail in \cref{sec:reporting_bug}.

\tool implements techniques based on natural language processing, dynamic software analysis, and automated bug report quality assessment. 
We designed and instantiated \tool as a stand-alone web system for Android apps that focuses on bugs manifesting in the app's UI.


We \textit{empirically evaluated} \tool, asking 18 end-users to report 12 bugs from six Android apps using \tool, and assess their experience. The reporters found \tool's guidance and automated suggestions/clarifications to be useful, accurate, and easy to use. Moreover, the bug reports collected by \tool are of higher quality than reports collected via a traditional template-based bug reporting system. 

\tool is an open-source tool hosted on GitHub~\cite{github} that can be used for any Android app project.  
More details about \ap's algorithms and evaluation can be found in its original research paper~\cite{Song:FSE22}.

\section{The \tool Interactive Bug Reporting Tool}

\tool is a standalone web-based chatbot, currently tailored for Android apps, that aims to collect high-quality information from the reporter through a guiding dialog. 
It generates a bug report with textual bug descriptions and app screen captures. 

\tool (i) guides the user in reporting essential bug report elements (the OB, EB, and S2Rs), (ii) checks the quality of these elements and offers instant feedback about issues, and (iii) provides graphical suggestions such as the next S2Rs.

\subsection{\tool's Graphical User Interface (GUI)}
\tool's GUI is composed of a standard chatbot interface and various panels for interactive bug reporting (see figure~1). 
The \textit{Chat Box} \img{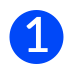} allows the reporter to provide textual descriptions of the OB, EB, and S2Rs and select \ap's graphical suggestions (\eg the next S2Rs via screenshots). 
The \textit{Reported Steps Panel} \img{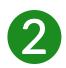} enumerates and displays the S2Rs that the user has reported, allowing them to edit the steps to correct mistakes. 
The \textit{Screen Capture Panel} \img{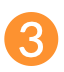} displays screen captures of the last three S2Rs. 
The \textit{Quick Action Panel}~\img{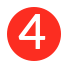} provides buttons to finish reporting the bug, restart the reporting session, and (pre)view the bug report. 
The \textit{Tips Panel}~\img{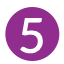} displays suggestions to reporters on how to use \ap and how to better express the OB, EB, and S2Rs. 
The tips change depending on the current stage of the conversation. 
\ap also provides a \textit{Developer Panel} that allows developers to add new apps to \ap (via the \textit{Settings} icon next to the \textit{Help} button).




 \begin{figure*}[t]
 	\centering{
 		\vspace{-0.5cm}
 		\includegraphics[width=0.9\linewidth]{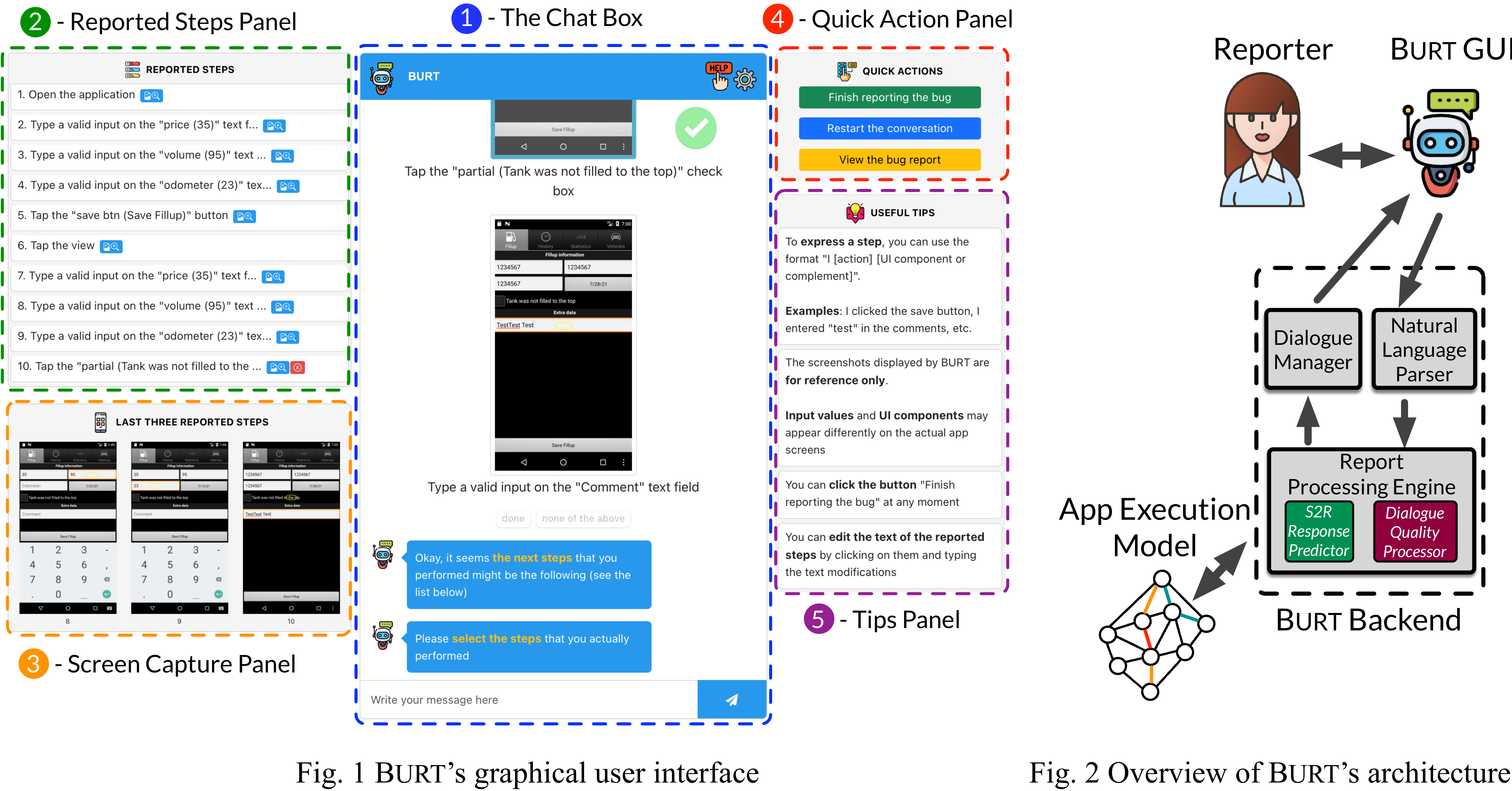}
 		\label{fig:burt}
	\vspace{-0.3cm}
 	}
 \end{figure*}

\subsection{Reporting a Bug with \tool}
\label{sec:reporting_bug}

To report a bug, the user first selects the target app exhibiting the defect by clicking on its icon---\tool lists the apps that it supports. 
Then, \ap guides the reporter through three phases: OB, EB, and S2R reporting. 
In each phase, \tool prompts the user to provide individual descriptions of the OB, EB, and S2Rs, respectively.  
\app automatically parses the descriptions and verifies their quality by matching them to states of a GUI-level execution model for the app (see Sec.~\ref{subsec:rp}). 

If the OB/EB/S2R is \textit{matched} to an 
app screen from \ap's execution model, \ap asks the user to confirm the matched screen. 
If the user confirms, \ap proceeds to the next phase of the conversation (\eg asking for the EB or next S2Rs); otherwise, \ap asks the user to rephrase the bug element.

If there are \textit{no} app screen matches, \ap suggests the user to revise their description and asks them to rephrase the OB/EB/S2R. 
With a new description, the quality verification is re-executed. 
If there are \textit{multiple} matches, \ap provides a list of up to five app screenshots (derived from the app execution model) that match the description.
The user can then inspect the app screens and select the one that she believes best matches her description of the bug element. 
If none are selected, \ap suggests additional app screens, if any. 
If the user selects one app screen, \ap saves the bug element description and screen, and proceeds to prompt the user for the next bug element. 
After three unsuccessful attempts to provide a high-quality description, \ap records the (last) provided description for bug report generation. 
This process proceeds for each bug element starting with the OB. 

\tool includes an additional feature to help users save time writing S2Rs: it suggests the probable next S2Rs that the user may have performed during actual app usage. 
\tool suggests the first five S2Rs from the most likely path from the current state to the OB state in the app execution model. 
This dialogue flow uses a predictive algorithm that uses \ap's execution model (see \cref{subsec:rp}).
The suggestions are displayed as a list of generated app screens, each screen representing a S2R. 
The generated screen is visually annotated with a yellow oval highlighting the GUI component (\eg a button) executed by the step. 
The user can select none, one, or multiple of the suggested S2Rs. 
When a S2R is selected, \ap suggests additional S2Rs, if any. 
When none are selected and \ap has more suggestions, \ap asks the user if they want additional suggestions. 
If so, \ap displays them. 
Otherwise, \ap prompts the user to describe the next S2R.

\subsection{Adding New Apps to \ap}

Developers can add new apps to \ap through the \textit{Developer Panel}, which requires the developer to upload the app icon and ZIP files containing app execution data required by \ap to build the app execution model. 
\ap uploads/extracts the files, parses the data to identify the app name and version, verifies the data format, and stores the data for later use. 
\ap provides feedback if there are any detected issues  with the data.
\looseness=-1

\vspace{-0.1cm}
\section{\tool's Architecture \& Implementation}
\label{sec:arch}


\tool has three main components (see figure~2). 
The \textit{Natural Language Parser (NL)} parses users' bug descriptions. 
The \textit{Dialogue Manager (DM)} implements the conversation flows for the reporting process. 
The \textit{Report Processing Engine (RP)} matches the parsed bug descriptions to the app execution model, to assess bug element quality and provide suggestions.

\subsection{Natural Language Parser (NL)}

\tool parses the textual OB/EB/S2R descriptions using dependency parsing via the Stanford CoreNLP toolkit~\cite{Manning2014}, which produces a tree of grammatical dependencies between words.
\ap first utilizes a heuristic-based approach from our prior work~\cite{Chaparro2017-2} to identify the sentence type of each user message (\eg conditional, imperative, or passive voice).  
Then, \tool executes one of its 16 parsing algorithms (one for each sentence type) that traverse the tree to extract the relevant words from the sentences. This parsing is based on our prior work on S2R quality assessment~\cite{Chaparro:FSE19} and extracts a phrase using the following format: \texttt{\small[subject] [action] [object] [preposition] [object2]}.   
For example, for the Mileage app~\cite{mileage}, the OB sentence \textit{``The fuel economy shows a NaN value on page''}, written in present tense, is parsed as \texttt{\small [fuel economy] [shows] [NaN value] [on] [page]}. 
The S2R sentence \textit{``Save the car fillup''}, written imperatively, is parsed as \texttt{\small [user] [saves] [car fillup]}. 

\subsection{Dialogue Manager (DM)}

\tool's dialogue flow guides users to report the OB, EB, and S2Rs. 
\ap's dialogue is multi-modal and capable of suggesting both natural language and graphical elements, \eg screenshots, to assist the user through the reporting process. 
The DM relies upon the RP engine to assess the quality of the bug elements reported by end users.
There are two main dialogue flows that \ap navigates: (i) performing quality checks on written bug report elements (applies to all bug elements), and (ii) automated suggestion of S2Rs (for S2Rs only). 

\subsection{Report Processing Engine (RP)}
\label{subsec:rp}

\noindent \tool's RP Engine consists of three sub-components:

(1) the \textit{App Execution Model} is a directed graph where nodes represent app screens and the edges are transitions between screens, triggered by GUI interactions (\eg \textit{taps} or \textit{type} events) on GUI components (\eg \textit{buttons} or \textit{text fields}). 
The graph screens contain the hierarchy of GUI components with their metadata (\eg labels) and a screen capture. The graph edges contain the action type, the interacted GUI component, and a screen capture highlighting the component. 
\ap builds the execution model from app usage data collected automatically, via automated app exploration, or manually (see \cref{subsec:exec-data}).
\looseness=-1

(2) the \textit{Dialogue Quality Processor} performs quality verification of the parsed OB/EB/S2R descriptions, by mapping them to app states/interactions from the execution model. 
A textual description is  \textit{high-quality} if it can be matched to the model, otherwise, it is \textit{low-quality}.  
This definition and \ap's dialogue features that prompt users to improve low-quality descriptions aim to reduce the knowledge gap between reporters and developers. 
To perform quality verification, \ap extends the bug description resolution/matching algorithm from our prior work~\cite{Chaparro:FSE19} and performs exploration of the execution model, driven by the matching of the reported OB/EB/S2Rs and user confirmations during the bug reporting process.

(3) the \textit{S2R Response Predictor} determines and suggests to the reporter the next S2Rs that she may have performed in practice. \ap implements a shortest-path approach to predict the next S2Rs~\cite{Song:FSE22}. 
\ap determines the paths between the current graph state and the corresponding OB state, and then, it computes the likelihood score based on the execution model edge weights, which are higher for manual app usage~\cite{Song:FSE22}. 

\subsection{Collecting App Execution Data}
\label{subsec:exec-data}

\tool requires app execution data for building the execution model. 
This data encodes sequential interactions made on the app features and comes from two sources: systematic app exploration via \crashscope~\cite{Moran:2017,7515457} and crowdsourced app usage (\eg from app users or developers). 
 
\crashscope's \textit{GUI-ripping engine} automatically generates app execution data 
(app interactions such as \textit{taps} or \textit{type} events) by running an app on a mobile device/emulator and a set of systematic exploration strategies (\eg top-down or bottom-up) on the app~\cite{Moran:2017}, to uncover as many app screens as possible and interact with most screen GUI components. 

Crowdsourced app usage data \textit{complements} the automated app usage data \cite{7180072}. 
This data can be collected through built-in trace recording app features that capture app usages ``in the wild'', or during in-house GUI-level app testing performed by developers. 
\app provides two tools to assist trace capture: \avt and \tracereplayer. 
\avt is a desktop app that allows humans to collect video recordings and \texttt{\small getevent} traces from a mobile device/emulator while humans are using an app.  \tracereplayer parses the collected \avt traces (\ie from humans) and converts them into the \crashscope data format. In this way, \tool can read/use this data to augment the app execution model (see details in \tool's repository~\cite{github}).

\subsection{\tool Implementation}

\ap is currently implemented as a web application, using the React Chatbot Kit~\cite{ReactChatbotKit} and Spring Boot~\cite{SpringBoot}. 
\tool also provides command-line tools for \crashscope and \tracereplayer, and the desktop \avt app, along with detailed documentation on how to use them. \ap's implementation is tailored for Android applications, however, its underlying techniques are generic enough to be easily adapted for other types of software---the App Execution Model Data Collection is the only platform-specific part.


\section{\app's Evaluation}

We conducted an empirical study to evaluate: (1) \ap's perceived  usefulness/usability (\textbf{RQ$_1$}/\textbf{RQ$_2$}); 
(2) \ap's intrinsic accuracy in performing bug report element quality verification and prediction (\textbf{RQ$_3$}); 
and (3) the quality of the bug reports collected with \ap  (\textbf{RQ$_4$}). 

\subsection{Methodology}
We selected 12 Android bugs (seven crashes, one handled error, and four non-crashes) from the dataset of our prior work~\cite{Cooper:ICSE21}. 
The bugs come from six Android apps of different domains: AntennaPod, Time Tracker, GnuCash, GrowTracker, and Droid Weight. 
To collect app execution data, we executed \crashscope on these apps and asked two computer science~(CS) students to use their main features---see our original paper for more details~\cite{Song:FSE22}. 
We recruited 18 participants to report these bugs (each reported three bugs) using \tool (most had little or no bug-reporting experience) and asked them to evaluate their experience via a questionnaire (with Likert-scale and open-ended questions). 
We analyzed the conversations the reporters had with \ap and measured how accurate \tool was during the reporting process. 
We asked additional 18 participants to report the same bugs with a template-based bug-reporting system (\aka~\itrac) and analyzed the collected bug reports to measure their quality based on the bug element correctness framework from our prior work~\cite{Chaparro:FSE19}. 
\itrac resembles existing issue trackers as it implements a web form with text fields and templates that explicitly ask for the bug summary/title and the OB/EB/S2Rs.
\looseness=-1


\subsection{Results}

\subsubsection{RQ$_1$/RQ$_2$: \app's User Experience}

The participants evaluated the \textbf{usefulness} of \tool's main features.
\vspace{0.3em}

\noindent\textit{\uline{Screen Suggestions}}: Half of the 18 reporters agreed that the screen suggestions were \textit{useful}, and another half (nine reporters) agreed that they were \textit{sometimes useful}. 
\vspace{0.3em}

\noindent\textit{\uline{OB/EB/S2R Quality Verification}}: The reporters had a positive impression on how often \tool correctly verified the quality of their bug descriptions. \tool was able to always (sometimes) verify the OB/EB/S2R descriptions of 9/10/11 (9/6/6) reporters (out of 18). Only two/one participant(s) felt that \tool rarely recognized and verified their EB/S2Rs. 
\vspace{0.3em}

\noindent\textit{\uline{B{\small URT} Messages \& Questions}}: 11 of 18 users often understood \ap's messages/questions, while 6 reporters understood them sometimes. Only one reporter rarely understood~them. 
\vspace{0.3em}

\noindent\textit{\uline{Panel of Reported S2Rs}}: \tool's panel of reported S2Rs was deemed to be useful by nine participants and somewhat useful by six participants -- one reporter found it somewhat useless.

\vspace{0.3em}

The reporters also assessed \tool's overall \textbf{ease of use}. 12 of 18 reporters indicated \tool was either easy or somewhat easy to use. Four reporters were neutral, while two reporters expressed that \app was somewhat difficult to use. 
\vspace{0.3em}

\textbf{RQ$_1$/RQ$_2$ Summary:} Overall, reporters found \ap's screen suggestions and S2R panel useful and \tool is easy to use.
Reporters also suggested improvements to \ap to support additional wording of bug report elements and provide more accurate suggestions. Improvements are planned for future work to improve \tool’s ability to recognize additional vocabulary and ways of phrasing the OB/EB/S2Rs.
\vspace{0.3em}

\subsubsection{RQ$_3$: \tool's Intrinsic Accuracy}

We analyzed the 54 conversations with \tool to assess how often \tool was able to 1) match OB/EB/S2R descriptions to the execution model, 
and 2) suggest relevant OB/S2R app screens to the reporters. 
\vspace{0.3em}

\noindent\textit{\uline{OB reporting}}: \burt matched the OB description to the correct screen in 3 of 54 (5.5\%) conversations and multiple screens in 35 of 54 (64.8\%) conversations. 
In 29 of 35 (80\%) conversations, the reporters selected one of the suggested screens. 
Overall, \burt correctly matched the users' OB descriptions in 32 of 54 (59.3\%) conversations. 

\vspace{0.3em}

\noindent\textit{\uline{EB reporting}}: \tool correctly matched the EB against the OB screen in 17 of 32 (53.1\%) cases
without having to ask the reporter for confirmation. 
In 6 of 32 (18.8\%) cases, \tool needed to ask the users for confirmation. 
In the remaining 9 cases, \tool struggled to parse the EB description.
\vspace{0.3em}

\noindent\textit{\uline{S2R reporting}}: In the 54 conversations, \tool matched 205 of the written S2R descriptions from the reporters. 
\tool matched the correct screen in 157 of 205 (76.6\%) cases. 
As for S2R prediction, among 32 conversations with a matched OB screen, S2R prediction occurred 146 times (mean: 4.6 per conversation). 
The reporters selected 1.6 of the 3.9 suggested S2Rs (on avg.) in 91 of 146 cases (62.3\%). 
\vspace{0.3em}

\textbf{RQ$_3$ Summary:} The results confirm the users' perception on the usefulness of screen suggestions and ability of performing correct bug element verification, which indicates that the techniques used in designing \ap's components are adequate.

\vspace{0.3em}

\subsubsection{RQ$_4$: Bug Report Quality}
We compared the quality of the $54\times2=108$ bug reports, collected with \itr and \ap. 

\vspace{0.3em}

\noindent\textit{\uline{S2R Quality}}: On average, \tool's reports contain fewer incorrect S2Rs than the \itrac reports (8.3\% vs. 20.4\%) and fewer missing S2Rs (19.4\% vs. 32\%). 
\vspace{0.3em}

\noindent\textit{\uline{OB/EB Quality}}. \tool and \itrac reports have a comparable number of incorrect EB descriptions (8 vs. 6 out of 54 reports). 
However, \tool reports have more incorrect OB descriptions compared to \itrac reports (16 vs. 8 out of 54 reports).  

\textbf{{RQ$_4$ Summary:}} \tool bug reports have higher-quality S2Rs than \itrac reports, and comparable EB descriptions. \app improvements are needed to better collect OB descriptions.

\vspace{-0.1cm}
\section{Most Related Existing Tools}

Moran \etal~\cite{Moran2015,7883352} proposed \fusion, a web system that allows the user to report the S2Rs graphically via dropdown lists of GUI components and actions (taps, swipes, \etc). Song \etal~\cite{Song:FSE20} proposed \bee, a GitHub plugin that identifies the type of each sentence in bug reports and alerts reporters of missing OB/EB/S2Rs. Fazzini~\etal proposed \ebug~\cite{Fazzini:TSE22}, a mobile app bug reporting system, similar to \fusion, that suggests possible future S2Rs as they are written. 
Shi \etal~\cite{shi2022buglistener} proposed \BugListener, an approach that identifies bug report dialogs in community live chats and synthesizes bug reports from them.
\looseness=-1

\ap has two main advancements over these prior tools: (1) it supports end-users with little or no bug reporting experience; for example, \fusion was not specifically designed to end-users, as inexperienced users found it \textit{more difficult} to use;
and (ii)
it offers an interactive interface, supporting automated suggestions, instant quality verification, and prompts for information clarification. 

\section{Final Remarks and Future Work}

\ap is a web-based chatbot for interactive bug reporting. 
Unlike existing bug reporting systems, \ap can guide end-users in reporting essential bug report elements (\ie OB, EB, and S2Rs), provide instant feedback about issues, and produce graphical suggestions of the elements that are likely to be reported next. 
\tool can help end-users easily report bugs and provide higher-quality bug reports. 
As future work, we plan to integrate \app with existing issue trackers.\looseness=-1



\section*{acknowledgements}

This work is supported in part by the NSF grants: CCF-1955837, CCF-1955853, CCF-2217733, and CCF-2007246. Any opinions, findings, and conclusions expressed herein are the authors and do not necessarily reflect those of the sponsors.

\balance
\bibliographystyle{IEEEtran}
\bibliography{references}

\end{document}